\documentstyle[12pt]{article}

\begin{document}
\thispagestyle{empty}
\font\fortssbx=cmssbx10 scaled \magstep2
\hbox to \hsize{
\hskip.5in \raise.1in\hbox{\fortssbx University of Wisconsin - Madison}
\hfill\vbox{\hbox{\bf MAD/PH/770}
            \hbox{June 1993}} }
\vspace{1in}

\begin{center}
{\Large\bf High Energy Neutrino Telescopes Detect Supernovae}\\
\vspace{1cm}
{\large F.~Halzen, J.~E.~Jacobsen}\\
{\it Department of Physics, University of Wisconsin, Madison  WI 53706}\\
\vspace{0.5cm}
{\large E.~Zas }\\
{\it Departamento de F\'\i sica de Part\'\i
culas, Universidad de Santiago\\
E-15706 Santiago de Compostela, Spain }
\end{center}

\vspace{1in}

\begin{abstract}

We have simulated the response of a high energy neutrino telescope to the
stream of low energy neutrinos produced by a supernova. The nominal threshold
of such detectors is in the GeV energy range. The passage of a large flux of
MeV neutrinos during a period of seconds will nevertheless be detected as an
excess of single counting rates in all individual optical modules. Detectors
under construction, which consist of roughly 200 modules, will be able to
detect a galactic supernova at or above the 5 $\sigma$ level. The rate of fake
signals is, however, too large for the telescope to serve as a neutrino watch.
Such capability requires detectors with roughly 3 times the number of optical
modules, thus within easy reach of the next generation detectors.

\end{abstract}

\newpage

The neutrino events detected in the Kamiokande\cite{Kamioka}, IMB\cite{IMB},
Baksan\cite{Baksan} and LSD\cite{LSD} detectors prior to the optical discovery
of supernova 1987A represented a most remarkable birth of neutrino astronomy.
Despite observation by four experiments, the data has left us with a variety
of questions and uncertainties. Most prominent is our inability to understand
the time of the Mont Blanc neutrino burst\cite{LSD} and the directionality of
the IMB events\cite{IMBang}. This underscores the importance of collecting as
much information as possible when presented with the rare opportunity of
observing the next nearby supernova. We here show that high energy neutrino
detectors, presently under construction\cite{Learned}, can observe the
neutrino bursts from galactic supernovae even though they have nominal
thresholds of GeV energy, i.e.\ almost three orders of magnitude above the MeV
energy of supernova neutrinos. Simultaneous detection of supernova neutrinos,
together with other devoted experiments\cite{Learned}, will help overall
statistics. Good relative timing between different experiments may allow the
determination of the location of the supernova if it is shielded from us by
large amounts of matter.

First generation high energy neutrino telescopes consist of approximately 200
optical modules (OM) deployed in deep, clear water or ice shielded from cosmic
rays. Coincident signals between the OMs detect the \v Cerenkov light of muons
with energy in excess of a few GeV. Also electromagnetic showers initiated by
very high energy electron neutrinos are efficiently detected. The idea has
been debated for some time whether these instruments have the capability to
detect the neutrinos from a supernova despite the fact that they are in the MeV
range. The production of copious numbers of positrons of tens of MeV energy
in the interaction of $\bar\nu_e$ with hydrogen, will suddenly yield signals
in all OMs for the 10 seconds duration of the burst. Clearly such a signal, no
matter how weak, will become statistically significant for a sufficient number
of OMs. We here perform a complete simulation of the signal and its detection
and conclude that the 200 OMs of detectors such as DUMAND and AMANDA are
sufficient to establish the occurence of a neutrino burst in coincidence with
the optical display of a supernova. We also show that the same detectors can
actually serve as a supernova watch, i.e.\ a ``fake" signal occurs less than
once a century, by increasing the number of OMs by a factor of three. This is
much less than the roughly 7000~OMs which are projected for a next-generation
detector\cite{FHVenice}.

Although aspects of the observations of SN1987A left some lingering doubts
about supernova models\cite{LSD,IMBang}, they remarkably confirmed the
established ideas for the supernova mechanisms\cite{mechanism}. At collapse the
core is expected to release energy in a prompt $\nu_e$ burst lasting a few
milliseconds. Most of the energy is however liberated after deleptonization in
a burst lasting about ten seconds. Roughly equal energies are carried by each
neutrino species. The time scale corresponds to the thermalization of the
neutrinosphere and its diffusion within the dense core\cite{sneutrinos}. Since
the ${\bar{\nu}}_e$ cross-section\cite{xsection} for the inverse beta decay
reaction on protons exceeds the characteristic cross sections for the other
neutrino flavors, ${\bar{\nu}}_e$ events dominate by a large factor after
including detection efficiency. In this reaction free protons absorb the
antineutrino to produce a neutron and a positron which is isotropically emitted
with an energy close to that of the initial neutrino. For the purpose of this
letter we use typical parameters, derived from SN1987 observations, which are
consistent with those previously estimated in supernova models. From the energy
distributions of the observed events the average temperature of the neutrino
sphere in SN1987A was deduced to be $4.0$~MeV\cite{mechanism}.

Before discussing our detailed Monte Carlo simulation we present a
back-of-the-envelope derivation of our final result. After convoluting the
4~MeV thermal Fermi distribution of the neutrinos with a detection cross
section rising with the square of the neutrino energy, one obtains an event
distribution peaked in the vicinity of 20~MeV. The tracklength of a 20~MeV
positron is roughly 10 centimeters and therefore over 3000 \v Cerenkov photons
are produced. This number combined with a typical quantum efficiency of
$25~\%$ leaves 800 detected photons in each event. Assuming these are emitted
within $\pi$ steradians, i.e.\ roughly the size of the \v Cerenkov cone, the
detection probability becomes a simple function of both the module collection
area $A_{M}$ and the distance to the positron shower $R$:
\begin{equation}
P(R) = min~\left[ { 800~A_{M} \over \pi~R^2},1 \right]
\end{equation}
For $R^2 < R_d^2~(\simeq 250~A_{M})$ the phototube will most likely trigger on
the positron while for larger $R$ the probability diminishes rapidly. By
evaluating the effective volume within a cone of $\pi$ steradians with vertex
at each OM, we obtain an approximate expression for the detection volume
associated with OMs:
\begin{eqnarray}
V_{eff} &\sim& {1 \over 4} \int_0^{R_{att}} P (R) R^2 dR\nonumber\\
& \sim & {1 \over 4} {\pi \over 3} R_d^3 + {1 \over 4} \int_{R_d}^{R_{att}}
800 A_{M} dR \nonumber\\
&\sim&  {\pi \over 12} R_d^3 + 200 A_{M} (R_{att}-R_d).
\end{eqnarray}
Here we have integrated up to the attenuation length $R_{att}$ of the medium
and divided by a factor of 4 in order to average over all orientations of the
\v Cerenkov cone. We conservatively assumed that the OM has $2\pi$ acceptance.
It is interesting to note that the effective volume is proportional to both
the collection area of the OM and the attenuation length. It is quite
insensitive to the solid angle over which the \v Cerenkov photons are
distributed ($\pi$). For OMs such as those used in the AMANDA detector with
collecting area $A_{M}=0.028$~m$^2$ and an attenuation length $R_{att}=25$~m
typical of ice\cite{iceatt}, we  obtain $V_{eff} \sim 130$~m$^3$. This result
can be used to rescale SN1987 observations to a supernova at a distance
$d_{kpc}$. From 11 events observed in 2.14~kton Kamiokande detector we
predict:
\begin{equation}
N_{Events} \sim 11~N_{M}~\left[{\rho~V_{eff} \over 2.14~kton} \right]
\left[ {52~kpc \over d_{kpc}} \right]^2
\end{equation}
for a detector with $N_{M}$ optical modules. For a 130~m$^3$ effective
volume of each of the 200 OMs we obtain 5300 events.

We now require a meaningful detection of this signal in the presence of the
continuous background counting rate of all phototubes. Over the 10~s duration
of the delayed neutrino burst from a supernova, the rms fluctuations of the
combined noise from all the OMs is:
\begin{equation}
\sigma_{1p.e.} = {\sqrt {10~\nu_{1p.e.}~N_{M}}}
\end{equation}
where the background counting rate in each module at the 1 photoelectron level
is represented by $\nu_{1p.e.}$. The probability that the noise in the OMs
fakes a supernova signal can be estimated assuming Poisson statistics. The
expected rate of supernova explosions in our galaxy is about
$2\times10^{-2}$~y$^{-1}$. If the detector is to perform a supernova
watch we must require that the frequency of fake signals is well below this
rate. The signal should therefore exceed $n_{\sigma} \ge 6$ which
corresponds to a probability of $9.9\times10^{-10}$. The corresponding number
of 10~seconds intervals indeed exceeds a century. Clearly the requirement can
be relaxed if we just demand that the detector can make a measurement in the
presence of independent confirmation. For an average noise rate of 1~kHz, a
typical value for the OMs in AMANDA, the rms fluctuation of the 20 million hits
expected in an interval of ten seconds is 1400. This implies that detection of
a galactic supernova is near the 4~$\sigma$ level for the 200 module
configuration, while detection should not represent a problem for the next
generation detector which consists of 7,000 OMs. Since the signal in the
present detector is marginal, it is necessary to do a more realistic
calculation of the event rate. We will conclude that our rough estimate is
somewhat conservative.

Background noise in the modules clearly plays a critical role so that low
noise environments such as ice have an intrinsic advantage. Furthermore, the
noise is expected to be reduced drastically at the 2 photoelectron level. This
will unfortunately also imply a reduction in effective volume for event
detection as we will see further on. Signal to noise is proportional to the
ratio $V_{eff}/{\sqrt{\nu}}$. Obviously increased attenuation length in the
medium and larger effective area of the OM results in an enhanced effective
volume. Considering parameters appropriate for DUMAND, an attenuation length
of 40~m in water and OMs with double diameter, we expect a factor 10 increase
in effective volume per optical module. This should readily compensate for a
noise rate higher by a factor 100. We therefore expect DUMAND and AMANDA to
have comparable sensitivity as supernova detectors.

For a complete calculation we have combined a detailed electromagnetic shower
Monte Carlo, initially developed to evaluate the radio emission by cascades in
ice\cite{zasemp}, with the AMANDA Monte Carlo. The shower Monte Carlo is a
fast three dimensional routine which simulates the dominant low energy
processes: M\o ller, Bhabha and Compton scattering as well as
electron-positron annihilation, continuous energy loss and multiple elastic
scattering as well as the bremsstrahlung and pair production processes which
dominate at high energy. We added the  capability to simulate the emission of
\v Cerenkov photons by cascade particles. Our event file consists of over
10,000 events sampled from a 4~MeV temperature Fermi-Dirac neutrino
distribution weighed by a cross section rising with the square of the neutrino
energy. Photon detection is simulated using the AMANDA Monte Carlo to
correctly account for the effects of attenuation in deep polar ice, optical
module efficiency as a function of photon wavelength and detector
geometry\cite{Amanda_MC}.

The final result can be quoted as an effective volume for the entire detector
of 23,000~m$^3 \pm 2,000$ for the first stage AMANDA configuration of 10
strings arranged in a nine 30~m side polygon with one at the center. Each
string has 20 modules spaced at 10~m intervals\cite{Amanda9+1}. We therefore
obtain an effective volume of 115~m$^3$ per module, close to our crude
estimate. Single OMs have a lower threshold than the Kamiokande and IMB
experiments which reconstruct \v Cerenkov cones. This is taken into account by
correction factors multiplying the event rates to be entered in Eq.~(3), which
we evaluated to be 1.4 (5.4) for Kamiokande (IMB). We thus obtain 7700 (9070)
events in the 200 modules of the 9+1 configuration of AMANDA. A trigger could
be implemented by monitoring the sum of all singles rates in $\sim 1$~s.
intervals, and requiring positive fluctuations in this sum for several such
intervals.

At the 2~p.e. level we obtain an effective volume of 5,900~m$^3 \pm 1,400$. If
the OM's noise rate is reduced by over a factor 16, signal-to-noise will
actually be improved by working at the 2~p.e. level. Clearly the statistical
significance of the signal will be improved by including information from
higher level triggers.

Typical noise rate for AMANDA modules imply that $\sigma_{1p.e.} \sim 1400$.
The calculated event rates for a supernova bursts in the center of the galaxy
therefore correspond to a 5.4 (6.4) sigma effect when rescaling the observed
signals from SN1987A at  Kamiokande (IMB). This should provide a sufficiently
clean signal. We can combine Eqs.~(2) and (3), including the correction
factor, to obtain the expected signal:
\begin{equation}
n_{\sigma}=0.35~ {\sqrt{N_{M}}}~
\left[{10^2 \over {\sqrt {10~\nu_{1p.e.}}}}\right]~
\left[{V_{eff} \over 125~m^3} \right]~
\left[ {8~kpc \over d_{kpc}} \right]^2~
\end{equation}
We here scaled to the Kamiokande event number. The constant is 0.42 for IMB.
The results are encouraging since a AMANDA supernova watch can be performed at
the 6-sigma level with only 340 (240) optical modules according to Kamiokande
(IMB) observations of SN1987A. A next-generation detector with over 7,000
modules should provide a sharp signal for a supernova at the galactic center
and should operate as a supernova watch to twice the distance to the galactic
center, i.e. covering all the galactic disk.

We end with a warning about the statistics. Clearly in a realistic analysis
penalty factors will be associated with various trials made to identify a
burst. This will, however, not alter our positive conclusions. A modest
increase the number of OMs can absorb the effect of a large number of trials,
e.g.\ associated with a sliding window to identify the 10~seconds burst.

\section*{Acknowledgements}

We are grateful to J. Hauptman, R.~Morse and J.~Learned for helpful
discussions. This work was supported in part by the University of Wisconsin
Research Committee with funds granted by the Wisconsin Alumni Research
Foundation, in part by the U.S.~Department of Energy under Contract
No.~DE-AC02-76ER00881, in part by the Texas National Research Laboratory
Commission under Grant No.~RGFY93-221 and in part by the ``Comit\'e Conjunto
Hispano-Norteamericano'' and CICYT.

\end{document}